\documentstyle[prb,aps,twocolumn,epsfig]{revtex}

\begin{document}

\title{Electrical spin injection and detection in a semiconductor. Is it feasible?}
\author{A. T. Filip, B.H. Hoving, F.J. Jedema, B.J. van Wees}
\address{Dept. of Applied Physics, Univ. of Groningen, Groningen 9747AG, The\\
Netherlands}
\author{B. Dutta, S. Borghs}
\address{IMEC, Kapeldreef 27, Leuven, Belgium}
\date{\today}
\maketitle

\begin{abstract}
The electrical injection of spin polarized electrons in a
semiconductor can  be achieved in principle by driving a current
from a ferromagnetic metal, where current is known to be
significantly spin polarized, into the semiconductor via ohmic
conduction. For detection a second ferromagnet can be used as
drain. We studied submicron lateral spin valve junctions, based on
high mobility InAs/AlSb two-dimensional electron gas (2DEG), with
Ni, Co and Permalloy as ferromagnetic electrodes. In the standard
geometry it is very difficult to separate true spin injection from
other effects, including local Hall effect, anomalous
magnetoresistance (AMR) contribution from the ferromagnetic
electrodes and weak localization/anti-localization corrections,
which can closely mimic the signal expected from spin valve
effect. The reduction in size, and the use of a multiterminal
non-local geometry allowed us to reduce the unwanted effects to a
minimum. Despite all our efforts, we have not been able to observe
spin injection. However, we find that this 'negative' result in
these systems is actually consistent with theoretical predictions
for spin transport in diffusive systems.
\end{abstract}

\pacs{PACS numbers: 72.20.My 85.80.Jm 73.61.Ey 72.15Gd}

% here starts the article

The idea to use the spin of the electron in electronic devices has
gained a lot of momentum lately, leading to the appearance of the
field of 'spintronics' \cite{kane}. It is envisioned that spin
sensitive electronics would open new perspectives to semiconductor
device technology. The potential to inject and control the
electronic spin in a semiconducting material is also of great
interest for the field of quantum computation \cite{diVincenzo}.
The first active device was suggested a decade ago by Datta and
Das \cite{datta}, who proposed an electronic device analogous to
the electro-optic modulator. The essential requirements for such a
device is the efficiency of injection of the spin polarized
carriers into the semiconductor and the long spin relaxation time.
The latter requirement was shown to be met in time resolved
optical experiments at low temperatures, where lifetimes as long
as $1 \mu s$ for spin in GaAs were observed \cite{awschalom1}.
Regarding the issue of spin injection, different approaches were
taken. Optical injection and detection of spin polarized carriers
in semiconductors have been shown in a experiment by J. Kikkawa
and D.D. Awschalom \cite{awschalom2}. Spin injection from a
ferromagnetic STM tip into GaAs has also been demonstrated
\cite{alvarado}. The electrical injection from a fully polarized
magnetic semiconductor, used as spin aligner, into a semiconductor
and optical detection was also shown \cite{schmidt}.

From  a device point of view, a major breakthrough would be to
have all electronic device, preferably operating at room
temperature. Therefore large efforts have been dedicated to
observe the spin valve effect, with semiconductors as the
'intermediate' layer \cite{cabbibo}. Recently Hammar {\it et.al.}
\cite{johnson1} have claimed the observation of electrical spin
injection in a 2DEG, by making use of the Rashba spin orbit
interaction in the semiconductor heterostructure as the detection
mechanism. However, this work has been commented upon and it was
suggested that in such a system the detection is not possible
within linear transport \cite{bart}, and the observed behavior is
probably related to a local Hall effect \cite{roukes}. Gardelis
{\it et.al.} \cite{gardelis} claim to have observed spin valve
effects in a semiconductor field effect transistor with Py source
and drain. A finite spin polarization of the semiconductor itself
was required in order to interpret the experimental observations
as spin valve effect. Another interesting approach has been taken
by Meier {\it et.al.} \cite{grundler}, who tried to observe spin
injection by modulating the spin orbit interaction via an external
gate. Hu {\it et.al.}  \cite{nitta}, by measuring in a
multi-injector HEMT geometry with ferromagnetic electrodes,
observed a gate and electrode spacing difference in the
magnetoresistive behavior, which they attributed to spin
injection.  However, the fact that the standard lateral spin valve
geometry leads to important local Hall phenomena has already been
pointed out \cite{tang}. Due to the dependence on the local
magnetization of the contacts, these spurious phenomena will often
closely resemble the signals expected from spin transport. In our
opinion, none of the previously mentioned experiments give an
unambiguous proof of spin dependent transport.

In our experiments, we considered the multi-terminal lateral spin
valve geometry, as depicted in fig. 1b. Two types of measurements
are possible. In the first one,  called the 'classic' spin valve
geometry, the current is injected and taken out from the
ferromagnetic electrodes. The voltage is measured between the same
electrodes, giving a standard four terminal measurement of the
junction. A second geometry, which we refer to as the non-local
geometry, corresponds to injecting current from the semiconducting
channel into the first ferromagnetic electrode  and measure the
voltage between the second ferromagnetic electrode and the
semiconducting channel (see fig. 1). Due to current polarization
in the injecting ferromagnet, at the interface a spin accumulation
will form, which will extend over a characteristic spatial
length-scale given by the spin flip length. If a second
ferromagnet is present in the vicinity of this interface, it can
be used as a spin sensitive voltage probe to detect this spin
accumulation. This is similar to the Johnson's potentiometric
method \cite{johnson2}, used for detecting spin accumulation in
Au. However, the essential advantage of a true lateral geometry
resides in the fact that no electrical current is flowing between
the injector and the detector electrodes. Therefore this geometry
allows to suppress any 'spin independent' magnetoresistive
contribution, i.e. the weak localization/anti-localization change
in conductivity of the semiconductor, and a possible
magnetoresistance contribution of the interface resistance.

\begin{figure}[hbp]
\centerline{\psfig{figure=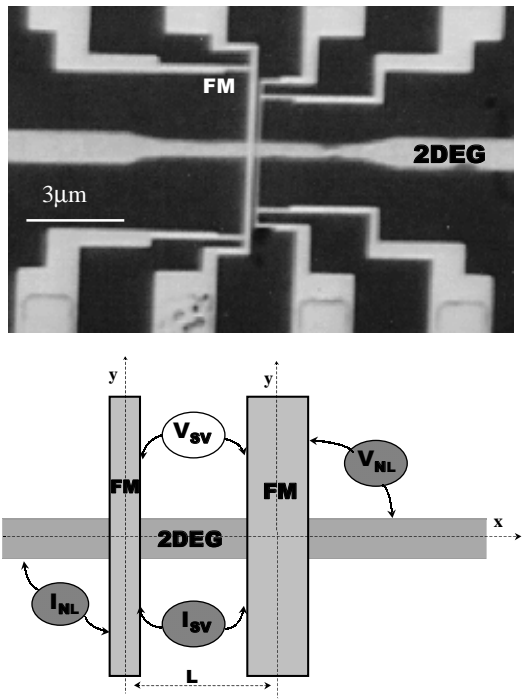,width=76mm}} \caption{ a)
SEM micrograph of  a sample. The $1\mu m$ wide 2DEG channel is
horizontal, and two ferromagnetic electrodes are vertical.  b)
Sketch of the two measurement configurations. The indices 'SV' and
'NL' refer to the classic spin-valve and, respectively, to the
non-local geometry. In the latter there is no current flow between
injector and detector.} \label{structure}
\end{figure}

The experiments were performed on devices made from high mobility
InAs/AlSb heterostructures, MBE grown on an GaAs substrate. Fig.1a
shows a SEM image of the FM/2DEG/FM junctions. Prior to
processing, the top barrier layer was removed by wet chemical
etching with Microposit MF321 photoresist developer. The exposed
15nm thick InAs layer hosts a 2DEG with an electron density
$n_{s}=1.5 \cdot 10^{16} m^{-2}$ and a mobility of $\mu = 1.5V/m^2
s$. In the first step 40nm thick Ti/Au metallization contacts were
deposited by means of optical lithography and e-beam evaporation.
An approx. $1 \mu m $ wide 2DEG channel was defined by optical
lithography and selective wet chemical etching, with a succinic
acid based solution. The use of wet etching techniques kept the
mesa at the smallest height possible, only 15nm. Consequently,
this allowed to reduce the spurious contribution due to local Hall
effects at the mesa edges to a minimum. In the last step the
ferromagnetic electrodes were defined by means of electron beam
lithography. In order to ensure different coercive fields the two
electrodes had different widths, 150 and $300nm$ in case of Py and
Co samples, and 150 and $450nm$ for the Ni samples. On all samples
the electrode lengths were 8 and respectively $12 \mu m$, the
spacing was $300nm$, and the thickness of the ferromagnetic layer
was $60nm$. Co and Py were deposited by sputtering, and Ni by
e-beam evaporation. Prior to deposition, the InAs surface was
cleaned by means of a low voltage Ar plasma etching. This was done
in order to remove the native InAs oxide and to ensure good ohmic
contact between the semiconductor and the ferromagnet. The
cleaning procedure is known to affect the InAs layer by enhancing
the electron density and reducing mobility. A a consequence, a
diffusive three dimensional InAs region is formed underneath the
ferromagnetic contacts. The square resistances were in the order
of $2-4\Omega$ for the ferromagnets and $300\Omega$ for the 2DEG
channel. The measured interface resistance were around $350\Omega$
and $750\Omega$ for the wide and, respectively, the narrow
electrode. Based on 2DEG material parameters, by evaluating the
number of modes in our channel, we calculated an average
ferromagnet/InAs interface transmission in the order of $30\%$.
For comparison, samples where the native InAs surface was left
intact were also made. In this case the contact resistance varied
between 10 and $100K\Omega$.

Measurements were performed by standard ac-lock-in techniques,
both at room temperature and at $4.2K$. The switching behavior of
the electrodes was characterized by four terminal anomalous
magnetoresistance (AMR) measurements of the ferromagnetic
electrodes. In most devices, in contrast to the room temperature
behavior, where a clear difference in the coercive fields of the
two electrodes could be established, the exact coercive fields at
helium temperature could not be inferred. At $4.2K$ the AMR curves
in parallel magnetic field showed only a smooth behavior,the
switching events being not visible. However, in some of the
devices clear switching of the magnetization direction of each
electrode could be observed.  Fig 2 shows one representative plot
of a Py/2DEG/Py device where the presence of different coercive
field for the two electrodes could be established. No resistance
modulation is observed when the two ferromagnets switch from a
parallel to an anti-parallel configuration. We carefully
characterized over 20 devices with different ferromagnetic
materials, out of which at least three showed switching events in
the $4.2K$ AMR curves, but no signal which could be attributed to
spin injection was observed.

\begin{figure}[phb]
\centerline{\psfig{figure=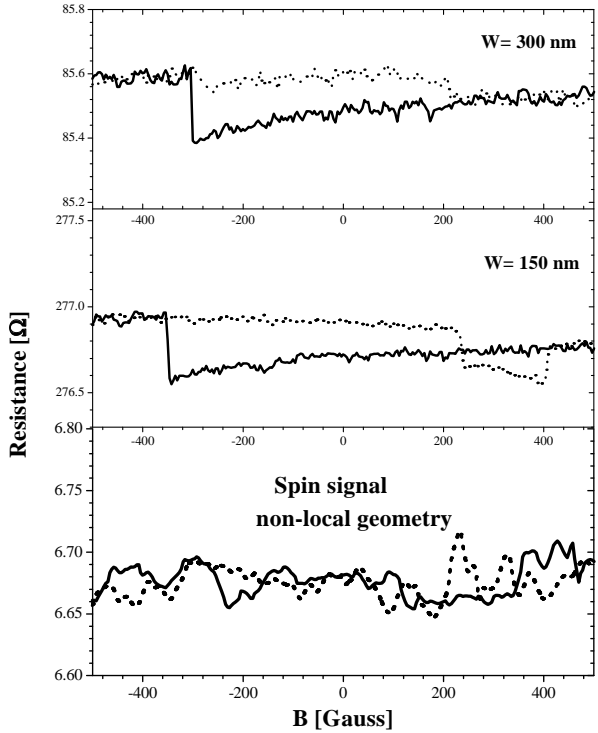,width=86mm}} \caption{ Spin
valve measurements for a Py/2DEG/Py device. Top two curves give
the AMR traces for the two  ferromagnetic electrodes, showing
different coercive fields in one sweep direction.No spin signal is
observed in any of the geometries. The dashed lines correspond to
a sweep of the magnetic field towards positive fields.}
\label{data}
\end{figure}

The outstanding question is to which extent can we understand
these results. Assuming weak spin scattering, the transport can be
described in terms of two independent spin channels. This
corresponds to an approach based on the standard Fert-Valet model
for describing spin transport \cite{fert}. The theoretical
implications for a two terminal geometry without spin flip
processes in the semiconductor have already been worked out by
Schmidt {\it et.al.} \cite{schmidt2}. Here we extend the analysis
to the multiterminal geometry sketched in fig. 1, and we also
allow for a finite spin flip length in the semiconductor. The
ferromagnets and the semiconductor are treated as diffusive 1D
channels. Therefore the transport properties of each channel are
fully determined by the bulk conductivities ($\sigma_F$ and,
respectively, $\sigma_N$), the spin flip lengths ($\lambda_F$ and
$ \lambda_N $), and, for the ferromagnet, the bulk spin
polarization of the current ( $ \alpha_F= \frac
 {\sigma_{\uparrow}-\sigma_{\downarrow}} {\sigma_{\uparrow} + \sigma{\downarrow}}
 $). If a  current is driven through such a non-homogeneous
system, the electrochemical potential for spin up and spin down
electrons ($\mu_{\uparrow}$ and $\mu_{\downarrow}$) can be
non-equal. This difference, due to different conductivities in the
two spin channels, will decay differently in a ferromagnet than in
a normal region, leading to a measurable voltage.

The spin transport, within the relaxation time approximation, is
described by the diffusion equation
\begin{equation}
 D  \frac { \partial^2 ( \mu_{\uparrow}-\mu_{\downarrow})}{\partial^2 x} =
\frac {( \mu_{\uparrow}-\mu_{\downarrow})}{\tau_{sf}}
 \end{equation}
where, $ \tau_{sf} $ is the spin-flip scattering time, and D is
the spin averaged diffusion constant
($D=({N_{\uparrow}+N_{\downarrow}})(N_{\uparrow}/D_{\uparrow}+N_{\downarrow}/D_{\downarrow})^{-1}$,
with $N(E_F)$ the density of states at the Fermi level). The
currents are related to electrochemical potentials via Ohm's law
 \begin{equation}
 j_{\uparrow, \downarrow}=-( \frac {\sigma_{\uparrow, \downarrow}}
 {e})
  \frac {\partial \mu_{\uparrow, \downarrow}} { \partial x}
 \end{equation}

The charge and spin conservation at each interface has also to be
taken into consideration.  We assume transparent interfaces, thus
we also require the equality of the chemical potential on both
sides of the interface.

By adding the appropriate boundary conditions at infinity, so that
far away from the interface one recovers the bulk transport
properties, the previous system of equations can be solved
analytically for the two geometries depicted in fig. 1a.

The resistance change between the parallel and the anti-parallel
configuration of the magnetizations of the two electrodes in the
'classic' spin valve geometry is given by
\begin{equation}
\Delta R_{SV}= 2R_{sq}\frac{\lambda_N}{w}
 \frac{{\alpha_F}^2} {(M^2+1)sinh(L/\lambda_N) + 2M
 cosh(L/\lambda_N)
 }
 \end{equation}
with

$M=1+\frac{\sigma_F}{\sigma_N}\frac{\lambda_N}{\lambda_F}(1-{\alpha_F}^2)$

$R_{sq}$ is the square resistance of the semiconductor, L is the
spacing between the two ferromagnets, and $w$ is the width of the
channel. In the non-local configuration the signal is reduced by a
factor of two $ \Delta R_{NL}=\frac{1}{2}\Delta R_{SV}$

In the limit $\lambda_N -> + \infty $ one recovers a result
similar to the one predicted by Schmidt {\it et.al.} for the
standard geometry \cite{schmidt2}

\begin{equation}
\Delta R_{NL} \approx R_{sq}\frac{{\lambda_F}^2}{w \cdot L}
 (\frac{{\sigma_F}} {{\sigma_N }})^2 \frac{{\alpha_F}^2}{1-{\alpha_F}^2}
 \end{equation}

The relevant range of parameters for  ferromagnet/2DEG/ferromagnet
junctions is $\sigma_F>>\sigma_N$ and $\lambda_N>>\lambda_F$,
meaning that, for a spin polarization of the ferromagnet smaller
than $100\%$, the conductivity mismatch correction factor M is
large, $M>>1$ . Then the expected signal can be expressed as

\begin{equation}
 \Delta R_{NL}= R_{sq}\frac{\lambda_N}{w} \frac{1}{sinh(L/ \lambda_N)} (\alpha_F/M)^2
\end{equation}
i.e. the injection efficiency is reduced from $\alpha_F$ to
$\alpha_F /M$. This shows spin valve signal is reduced due to the
conductivity mismatch between the semiconductor and the
ferromagnet. Moreover, the spin injection efficiency is very
sensitive to the spin flip length in the ferromagnetic material.
If this length is small, the expected spin signal is also reduced.

Based on  GMR experiments \cite{dubois}, a spin flip length
between 8 and  $40nm$ and a bulk current spin polarization around
$35\%$ is expected for Py. Assuming for the 2DEG a spin
flip-length an order of magnitude of $1 \mu m$, we obtain the
reduction in spin injection efficiency, $M=90$. This corresponds
to an absolute signal of  $0.2m\Omega$, or in the order of
magnitude of $10^{-6}$ of the square resistance. The best signal
resolution we could obtained was only $5m\Omega$, so the expected
spin signal was well below the sensitivity threshold.

The direct conclusion to be extracted from the modeling, also
pointed out by Schmidt {\it et.al.} \cite{schmidt2}, is that the
conductivity mismatch blocks spin injection. This result is
stemming from the fact that the lowest conductance in the problem,
the conductance of the semiconductor, is spin independent. One
possible solution is to make use of magnetic semiconductors, with
low conductivity or very high spin polarization, as in the
experiments of Fielderling {\it et.al.} and Ohno {\it et.al.}
\cite{schmidt}. A second choice would be to use tunnel barriers as
the injecting mechanism, where the spin polarization of the
tunneling current depends directly on the products of the
densities of states in the two materials.

One more aspect should also be considered: what is the actual
reliability of the model. Recently we were able to observe spin
valve effects in a similar geometry with Cu replacing the
semiconductor as the normal channel \cite{friso}. Using the values
obtained in GMR experiments for the spin flip lengths and spin
polarization in the ferromagnet \cite{dubois}, the order of
magnitude of the observed effect was in  quantitative agreement to
the theoretical predictions. Obviously, the main difference in the
all metal devices was the absence of conductivity mismatch between
the two materials. A potential limitation in the semiconductor
case is the fact that the 2DEG channel is quasi-ballistic.
Nevertheless, the presence of the diffusive regions underneath the
ferromagnetic contacts should allow us to use a diffusive model to
describe spin injection. Moreover, the conductivity mismatch
arguments should also be valid for a purely ballistic channel. In
that case, the expected signal should be given by an analogous of
eq. 5, with the diffusive 1D conductivity of the semiconductor
being replaced by the inverse of the Sharvin resistance, due to
presence of only a few model in the 2DEG channel. Thus the
conductivity mismatch arguments should be valid in any device with
where the intermediate region has the lowest conductivity, for
example in the case of carbon nanotubes \cite{alphenaar}.

In conclusion, submicron lateral spin  valve structures in high
mobility InAs/AlSb heterostructures have been fabricated, with Ni
Co and Py as ferromagnetic electrodes. Despite all efforts to
improve signal resolution and eliminate spurious effects, no spin
injection was observed. By no means this 'negative' outcome of our
experiments can be considered as a proof that spin injection in a
semiconductor is not possible with usual metallic ferromagnets.
However, the agreement with theoretical predictions casts some
doubt on the feasibility of straightforward spin injection from a
metallic ferromagnet into a semiconductor.

This work was supported by the Dutch Foundation for Fundamental
Research on Matter (FOM)and European Commission (ESPRIT-MELARI
consortium Spider). We acknowledge useful discussions with G.
Schmidt and L. Molenkamp. We thank T.M. Klapwijk for his
stimulating support in this work.

\end{document}